# Comments on «The Behavior of Some Thermodynamic Characteristics of Single-Component Substance in the Region Defined by the Line of Liquid–Vapor Equilibrium»

**I.H. Umirzakov**
*Institute of thermophysics, Siberian Division, Russian Academy of Sciences*
*Novosibirsk, 630090, Russia*
*e-mail: cluster125@gmail.com*

**Abstract -** It is shown that in general case may be not correct the statements of [1,2,6-8] that 1) the isochoric heat capacity on the entire thermodynamic surface, including the metastable region of states and the region defined by the spinodal, remains positive and finite except for the critical point, and 2) the isobaric heat capacity becomes negative in the region defined by spinodal.

Thermodynamic analysis in ref. [1,2] is performed of the behavior of a number of thermodynamic characteristics of single-component substance under conditions of hypothetical continuous liquid-vapor phase transition. According [1,2] «it is demonstrated that the isochoric heat capacity on the entire thermodynamic surface, including the metastable region of states and the region defined by the spinodal, remains positive and finite except for the critical point».

We will consider the behavior of a thermodynamic system under conditions of continuous variation of the parameters of state and assume that the thermodynamic surface of states does not experience shocks and discontinuities [1,2].

Using the equation [3]

$$C_P - C_V = -T \cdot (\partial p / \partial T)_V^2 / (\partial p / \partial V)_T , \qquad (1)$$

where $T$ is the temperature, $p$ is the pressure, $V$ is the volume, $C_P = (\partial H / \partial T)_P$ is the isobaric heat capacity, $C_V = (\partial U / \partial T)_V$ is the isochoric heat capacity, $H$ is the enthalpy, and $U$ is the internal energy of the thermodynamic system, one can conclude [1,2] that because

$$(\partial p / \partial V)_T < 0 \qquad (2)$$

everywhere outside of the spinodal, then

$$C_P - C_V > 0 , \qquad (3)$$

$$C_P > C_V . \qquad (4)$$

We have from (1)-(4) that, particularly,

$$\alpha \equiv C_V / C_P < 0, \quad \text{if} \quad 0 < C_P < \infty, \quad C_V < 0; \qquad (5)$$

$$\alpha < 0, \quad \quad \text{if} \quad -\infty < C_P < 0, \quad C_V > 0; \qquad (6)$$

$$\alpha = 0, \qquad \text{if } 0 < |C_P| < \infty, \quad C_V = 0; \tag{7}$$

$$\alpha = 0, \qquad \text{if } |C_P| = \infty, \quad |C_V| < \infty; \tag{8}$$

$$\alpha > 1, \qquad \text{if } -\infty < C_P < 0, \quad C_V < 0. \tag{9}$$

From (5)-(9) we conclude that the inequalities

$$0 < \alpha < 1 \tag{10}$$

of [1,2] (see (2) from [1,2]) is not correct in general case. One can easily see from (1)-(4) that the inequalities (10) are true, if

$$0 < C_P < \infty, \quad C_V > 0. \tag{11}$$

One can obtain from (1) the equation [1,2]

$$\alpha \equiv C_V / C_P = 1 - T \cdot (\partial p / \partial T)_V / (\partial H / \partial V)_P, \tag{12}$$

which can be represented as

$$(\partial H / \partial V)_P = T \cdot (\partial p / \partial T)_V / (1 - \alpha). \tag{13}$$

The inequalities

$$0 < (\partial p / \partial T)_V \neq \infty \tag{14}$$

are true everywhere on the thermodynamic surface [4].
From (13) and (14) we conclude that

$$0 < (\partial H / \partial V)_P \neq \infty \tag{15}$$

in the entire region between bimodal and spinodal, if the inequalities

$$-\infty \neq \alpha < 1 \tag{16}$$

are true. The inequalities (16) are true, if the inequalities (4)-(8) and (10)-(11) take place. From (2), (14) and (15) we conclude that

$$0 \leq C_P = -(\partial H / \partial V)_P \cdot (\partial p / \partial T)_V / (\partial p / \partial V)_T \neq \infty. \tag{17}$$

We conclude that $C_P = 0$ or $C_P > 0$ is not unknown and therefore the statement of [1,2] about positivity of $C_P$ is not correct, if (2), (4)-(8), (10)-(11), (14) and (15) are take place.
But $C_P > 0$ if

$$-\infty \neq (\partial p / \partial V)_T < 0, \tag{18}$$

therefore $0 < C_V = \alpha \cdot C_P \neq \infty$ only in the case when the inequalities (4)-(6), (10)-(11), (14) and (18) take place. So the statement of [1,2] that the isochoric heat capacity $C_V$ is positive and remains finite in the entire region between bimodal and spinodal is correct only in this case, and the statement is not proved in [1,2] under conditions (2) and (14).

There is no proof in [1,2] that the inequalities (7)-(9) and

$$(\partial p / \partial V)_T = -\infty \tag{19}$$

cannot take place. Therefore the statement (N1) that the isochoric heat capacity is positive and remains finite in the entire region between bimodal and spinodal is not proved in [1,2] under conditions (2) and (14).

So we proved that the statement (N2) of [1,2] that «the heat capacity $C_V$ remains a positive and finite quantity in the entire range from binodal to spinodal in both liquid and gas regions only if the compressibility remains finite and negative» is not correct.

If

$$\alpha^{sp} = 0 \tag{20}$$

on the spinodal, then from (13) and (14) we have

$$(\partial H / \partial V)_P^{sp} = T \cdot (\partial p / \partial T)_V^{sp}, \tag{21}$$

$$C_P^{sp} = -(\partial H / \partial V)_P^{sp} \cdot (\partial p / \partial T)_V^{sp} / (\partial p / \partial V)_T^{sp} = -T \cdot [(\partial p / \partial T)_V^2]^{sp} / (\partial p / \partial V)_T^{sp} = \pm \infty. \tag{22}$$

The equations (22) are true because when passing the spinodal, the derivative $(\partial p / \partial V)_T$ changes its value from (–0) to (+0) [1,2].

Using (13) we have

$$C_P = -(\partial H / \partial V)_P \cdot (\partial p / \partial T)_V / (\partial p / \partial V)_T = -T \cdot [(\partial p / \partial T)_V]^2 / (\partial p / \partial V)_T / (1 - \alpha). \tag{23}$$

If

$$(\partial p / \partial V)_T > 0 \tag{24}$$

everywhere in the region defined by spinodal, and (16) takes place, then the statement (N3) of [1,2] that the isobaric heat capacity becomes negative in this entire region defined by spinodal is correct. But (16) takes place if the inequalities (6-8) and (10)-(11) are take place.

There is no proof in [1,2] that the inequalities (9) or one of the equalities

$$(\partial p / \partial V)_T = 0, \quad (\partial p / \partial V)_T = \infty,$$

cannot take place. So there is no proof in [1,2] that the statement N3 is valid.

The exact equations

$$C_V = (\partial p / \partial T)_V \cdot [(\partial U / \partial V)_T - (\partial U / \partial V)_P] / (\partial p / \partial V)_T, \tag{25}$$

$$C_V = (\partial p / \partial T)_V \cdot [(\partial p / \partial T)_V - (\partial H / \partial V)_P] / (\partial p / \partial V)_T. \tag{26}$$

are valid [1,2,5].

$(\partial p/\partial V)_T = 0$

on the spinodal and the numerator in Eq. (26) on the spinodal is equal to zero according to (21), therefore an indeterminacy of the form 0/0 arises, which must be uncovered, for example, by the L'Hospital rule [1,2].

The positivity of $C_V$ outside of the spinodal and (25) were used in [1,2] in order to prove that $C_V$ is positive and finite in the region defined by spinodal. However we have shown above that there is no proof of the positivity of $C_V$ outside of the spinodal in [1,2]. Therefore there is no proof in [1,2] that the isochoric heat capacity has positive and finite values in the region defined by spinodal.

Therefore the statement (N4) of [1,2] that the condition of thermal stability $C_V > 0$ is not violated and is valid everywhere on the thermodynamic surface, including on the spinodal and in the region defined by spinodal on the $p-V$ diagram has no proof in [1,2].

The same and similar «results» as in [1,2] and the same «proofs» of the «results» as in [1,2] were obtained in [6-8]. From above consideration we conclude there are no proofs of «results» of [6-8] in [6-8].